
%

\newcount\driver \driver=1          

\newcount\mgnf\newcount\tipi\newcount\tipoformule
\mgnf=0          
\tipi=2          
\tipoformule=1   
\ifnum\mgnf=0
   \magnification=\magstep0\hoffset=0.cm
   \voffset=-0.5truecm\hsize=16.5truecm\vsize=24.truecm
   \parindent=4.pt\fi
\ifnum\mgnf=1
   \magnification=\magstep1\hoffset=0.truecm
   \voffset=-0.5truecm\hsize=16.5truecm\vsize=24.truecm
   \baselineskip=14pt plus0.1pt minus0.1pt \parindent=12pt
   \lineskip=4pt\lineskiplimit=0.1pt      \parskip=0.1pt plus1pt\fi
%
%
\let\a=\alpha   \let\g=\gamma    \let\d=\delta \let\e=\varepsilon
\let\z=\zeta     \let\th=\vartheta \let\l=\lambda
    \let\n=\nu    \let\x=\xi           
\let\s=\sigma \let\t=\tau    
\let\ps=\psi  \let\o=\omega


{\count255=\time\divide\count255 by 60 \xdef\oramin{\number\count255}
        \multiply\count255 by-60\advance\count255 by\time
   \xdef\oramin{\oramin:\ifnum\count255<10 0\fi\the\count255}}

\def\ora{\oramin }

\def\data{\number\day/\ifcase\month\or gennaio \or febbraio \or marzo \or
aprile \or maggio \or giugno \or luglio \or agosto \or settembre
\or ottobre \or novembre \or dicembre \fi/\number\year;\ \ora}

\setbox200\hbox{$\scriptscriptstyle \data $}

\newcount\pgn \pgn=1
\def\foglio{\number\numsec:\number\pgn
\global\advance\pgn by 1}
\def\foglioa{A\number\numsec:\number\pgn
\global\advance\pgn by 1}

%


\global\newcount\numsec\global\newcount\numfor
\global\newcount\numfig
\gdef\profonditastruttura{\dp\strutbox}
\def\senondefinito#1{\expandafter\ifx\csname#1\endcsname\relax}
\def\SIA #1,#2,#3 {\senondefinito{#1#2}
\expandafter\xdef\csname #1#2\endcsname{#3} \else
\write16{???? ma #1,#2 e' gia' stato definito !!!!} \fi}
\def\etichetta(#1){(\veroparagrafo.\veraformula)
\SIA e,#1,(\veroparagrafo.\veraformula)
 \global\advance\numfor by 1
 \write16{ EQ \equ(#1) == #1  }}
\def \FU(#1)#2{\SIA fu,#1,#2 }
\def\etichettaa(#1){(A\veroparagrafo.\veraformula)
 \SIA e,#1,(A\veroparagrafo.\veraformula)
 \global\advance\numfor by 1
 \write16{ EQ \equ(#1) == #1  }}
\def\getichetta(#1){Fig. \verafigura
 \SIA e,#1,{\verafigura}
 \global\advance\numfig by 1
 \write16{ Fig. \equ(#1) ha simbolo  #1  }}
\newdimen\gwidth
\def\BOZZA{
\def\alato(##1){
 {\vtop to \profonditastruttura{\baselineskip
 \profonditastruttura\vss
 \rlap{\kern-\hsize\kern-1.2truecm{$\scriptstyle##1$}}}}}
\def\galato(##1){ \gwidth=\hsize \divide\gwidth by 2
 {\vtop to \profonditastruttura{\baselineskip
 \profonditastruttura\vss
 \rlap{\kern-\gwidth\kern-1.2truecm{$\scriptstyle##1$}}}}}
\footline={\rlap{\hbox{\copy200}\ $\st[\number\pageno]$}\hss\tenrm
\foglio\hss}
}
\def\alato(#1){}
\def\galato(#1){}
\def\veroparagrafo{\number\numsec}\def\veraformula{\number\numfor}
\def\verafigura{\number\numfig}
\def\geq(#1){\getichetta(#1)\galato(#1)}
\def\Eq(#1){\eqno{\etichetta(#1)\alato(#1)}}
\def\eq(#1){\etichetta(#1)\alato(#1)}
\def\Eqa(#1){\eqno{\etichettaa(#1)\alato(#1)}}
\def\eqa(#1){\etichettaa(#1)\alato(#1)}
\def\eqv(#1){\senondefinito{fu#1}$\clubsuit$#1\write16{No translation for #1}%
\else\csname fu#1\endcsname\fi}
\def\equ(#1){\senondefinito{e#1}\eqv(#1)\else\csname e#1\endcsname\fi}

\ifnum\tipoformule=1\let\Eq=\eqno\def\eq{}\let\Eqa=\eqno\def\eqa{}
\def\equ{{}}\fi

\def\include#1{
\openin13=#1.aux \ifeof13 \relax \else
\input #1.aux \closein13 \fi}
\openin14=\jobname.aux \ifeof14 \relax \else
\input \jobname.aux \closein14 \fi
%
%
%
\newdimen\xshift \newdimen\xwidth
%
%
\def\ins#1#2#3{\vbox to0pt{\kern-#2 \hbox{\kern#1 #3}\vss}\nointerlineskip}
%
%
\def\insertplot#1#2#3#4{
    \par \xwidth=#1 \xshift=\hsize \advance\xshift
     by-\xwidth \divide\xshift by 2 \vbox{
  \line{} \hbox{ \hskip\xshift  \vbox to #2{\vfil
 \ifnum\driver=0 #3  
                 \special{ps: plotfile #4.ps} 
 \fi
 \ifnum\driver=1  #3    \includegraphics{#4.ps}       \fi
 \ifnum\driver=2  #3   \ifnum\mgnf=0
                       \special{#4.ps 1. 1. scale}\fi
                       \ifnum\mgnf=1
                       \special{#4.ps 1.2 1.2 scale}\fi\fi
 \ifnum\driver=3 \ifnum\mgnf=0
                 \psfig{figure=#4.ps,height=#2,width=#1,scale=1.}
                 \kern-\baselineskip #3\fi
                 \ifnum\mgnf=1
                 \psfig{figure=#4.ps,height=#2,width=#1,scale=1.2}
                 \kern-\baselineskip #3\fi
 \ifnum\driver=5  #3  \fi
\fi}
\hfil}}}

\newskip\ttglue
\def\TIPI{
\font\ottorm=cmr8   \font\ottoi=cmmi8
\font\ottosy=cmsy8  \font\ottobf=cmbx8
\font\ottott=cmtt8  
\font\ottoit=cmti8
\def \ottopunti{\def\rm{\fam0\ottorm}
\textfont0=\ottorm  \textfont1=\ottoi
\textfont2=\ottosy  \textfont3=\ottoit
\textfont4=\ottott
\textfont\itfam=\ottoit  \def\it{\fam\itfam\ottoit}%
\textfont\ttfam=\ottott  \def\tt{\fam\ttfam\ottott}%
\textfont\bffam=\ottobf
\normalbaselineskip=9pt\normalbaselines\rm}
\let\nota=\ottopunti}
\def\TIPIO{
\font\setterm=amr7 
\font\settesy=amsy7 \font\settebf=ambx7 
\def \settepunti{\def\rm{\fam0\setterm}
\textfont0=\setterm   
\textfont2=\settesy   
\textfont\bffam=\settebf  \def\bf{\fam\bffam\settebf}
\normalbaselineskip=9pt\normalbaselines\rm
}\let\nota=\settepunti}

\def\TIPITOT{
\font\twelverm=cmr12
\font\twelvei=cmmi12
\font\twelvesy=cmsy10 scaled\magstep1
\font\twelveex=cmex10 scaled\magstep1
\font\twelveit=cmti12
\font\twelvett=cmtt12
\font\twelvebf=cmbx12
\font\twelvesl=cmsl12
\font\ninerm=cmr9
\font\ninesy=cmsy9
\font\eightrm=cmr8
\font\eighti=cmmi8
\font\eightsy=cmsy8
\font\eightbf=cmbx8
\font\eighttt=cmtt8
\font\eightsl=cmsl8
\font\eightit=cmti8
\font\sixrm=cmr6
\font\sixbf=cmbx6
\font\sixi=cmmi6
\font\sixsy=cmsy6
\font\twelvetruecmr=cmr10 scaled\magstep1
\font\twelvetruecmsy=cmsy10 scaled\magstep1
\font\tentruecmr=cmr10
\font\tentruecmsy=cmsy10
\font\eighttruecmr=cmr8
\font\eighttruecmsy=cmsy8
\font\seventruecmr=cmr7
\font\seventruecmsy=cmsy7
\font\sixtruecmr=cmr6
\font\sixtruecmsy=cmsy6
\font\fivetruecmr=cmr5
\font\fivetruecmsy=cmsy5
\textfont\truecmr=\tentruecmr
\scriptfont\truecmr=\seventruecmr
\scriptscriptfont\truecmr=\fivetruecmr
\textfont\truecmsy=\tentruecmsy
\scriptfont\truecmsy=\seventruecmsy
\scriptscriptfont\truecmr=\fivetruecmr
\scriptscriptfont\truecmsy=\fivetruecmsy
\def \eightpoint{\def\rm{\fam0\eightrm}
\textfont0=\eightrm \scriptfont0=\sixrm \scriptscriptfont0=\fiverm
\textfont1=\eighti \scriptfont1=\sixi   \scriptscriptfont1=\fivei
\textfont2=\eightsy \scriptfont2=\sixsy   \scriptscriptfont2=\fivesy
\textfont3=\tenex \scriptfont3=\tenex   \scriptscriptfont3=\tenex
\textfont\itfam=\eightit  \def\it{\fam\itfam\eightit}%
\textfont\slfam=\eightsl  \def\sl{\fam\slfam\eightsl}%
\textfont\ttfam=\eighttt  \def\tt{\fam\ttfam\eighttt}%
\textfont\bffam=\eightbf  \scriptfont\bffam=\sixbf
\scriptscriptfont\bffam=\fivebf  \def\bf{\fam\bffam\eightbf}%
\tt \ttglue=.5em plus.25em minus.15em
\setbox\strutbox=\hbox{\vrule height7pt depth2pt width0pt}%
\normalbaselineskip=9pt
\let\sc=\sixrm  \let\big=\eightbig  \normalbaselines\rm
\textfont\truecmr=\eighttruecmr
\scriptfont\truecmr=\sixtruecmr
\scriptscriptfont\truecmr=\fivetruecmr
\textfont\truecmsy=\eighttruecmsy
\scriptfont\truecmsy=\sixtruecmsy
}\let\nota=\eightpoint}

\newfam\msbfam   
\newfam\truecmr  
\newfam\truecmsy 
\newskip\ttglue
\ifnum\tipi=0\TIPIO \else\ifnum\tipi=1 \TIPI\else \TIPITOT\fi\fi

\def\didascalia#1{\vbox{\nota\0#1\hfill}\vskip0.3truecm}

%
\def\V#1{\vec#1}
\def\T#1{#1\kern-4pt\lower9pt\hbox{$\widetilde{}$}\kern4pt{}}

\let\dpr=\partial\let\io=\infty
\def\fra#1#2{{#1\over#2}}
\let\0=\noindent

\def\guida{\leaders\hbox to 1em{\hss.\hss}\hfill}
\def\tende#1{\vtop{\ialign{##\crcr\rightarrowfill\crcr
              \noalign{\kern-1pt\nointerlineskip}
              \hglue3.pt${\scriptstyle #1}$\hglue3.pt\crcr}}}
\def\otto{{\kern-1.truept\leftarrow\kern-5.truept\to\kern-1.truept}}

\def\acapo{\hfill\break}

\let\ciao=\bye

\def\etc{\hbox{\it etc}}

\def\ie{\hbox{\it i.e.\ }}


\def\AA{{\V A}}\def\aa{{\V\a}}\def\nn{{\V\n}}\def\oo{{\V\o}}
\def\mm{{\V m}}\def\nn{{\V\n}}\def\lis#1{{\overline #1}}
\def\FF{{\cal F}}
\def\={{ \; \equiv \; }}

\def\Dpr{{\V \dpr}\,}
\def\Im{{\rm\,Im\,}}

\def\pps{{\V\ps{\,}}}

\def\2{{1\over2}}

\def\tst{\textstyle}
\def\st{\scriptscriptstyle}
\let\\=\noindent

\def\*{\vskip0.3truecm}

\catcode`\%=12\catcode`\}=12\catcode`\{=12
              \catcode`\<=1\catcode`\>=2

\openout13=gvnn.ps
\write13<
\write13<
\write13</punto { gsave 2 0 360 newpath arc fill stroke grestore} def>
\write13<0 90 punto     >
\write13<70 90 punto    >
\write13<120 60 punto   >
\write13<160 130 punto  >
\write13<200 110 punto  >
\write13<240 170 punto  >
\write13<240 130 punto  >
\write13<240 90 punto   >
\write13<240 0 punto    >
\write13<240 30 punto   >
\write13<210 70 punto   >
\write13<240 70 punto   >
\write13<240 50 punto   >
\write13<0 90 moveto 70 90 lineto>
\write13<70 90 moveto 120 60 lineto>
\write13<70 90 moveto 160 130 lineto>
\write13<160 130 moveto 200 110 lineto>
\write13<160 130 moveto 240 170 lineto>
\write13<200 110 moveto 240 130 lineto>
\write13<200 110 moveto 240 90 lineto>
\write13<120 60 moveto 240 0 lineto>
\write13<120 60 moveto 240 30 lineto>
\write13<120 60 moveto 210 70 lineto>
\write13<210 70 moveto 240 70 lineto>
\write13<210 70 moveto 240 50 lineto>
\write13<stroke>
\closeout13
\catcode`\%=14\catcode`\{=1
\catcode`\}=2\catcode`\<=12\catcode`\>=12


\vglue0.truecm


\0{\bf Twistless KAM tori.}\footnote{${}^*$}{\nota Archived in
{\tt mp\_arc@math.utexas.edu}; to get a TeX version, send an empty
E-mail message.}

\vskip1.truecm
\0{\bf Giovanni Gallavotti\acapo
\rm CNRS-CPT Luminy, case 907\acapo
F-13288 Marseille}\footnote{${}^1$}{\nota
E-mail: {\tt gallavotti\%40221.hepnet@lbl.gov}; permanent address:
Dipartimento di Fisica,
Universit\`a di Roma, ``La Sa\-pi\-en\-za", P. Moro 2, 00185 Roma,
Italia.}
\vskip0.5truecm
\0{\bf Abstract:} {\sl A selfcontained proof of the KAM theorem in the
Thirring model is discussed.}
\vskip0.5truecm
\0{\sl Keywords:\it\ KAM, invariant tori, classical mechanics, perturbation
theory, chaos}
\vskip0.5truecm
\numsec=1\numfor=1\pgn=1
I shall particularize the Eliasson method, [E], for KAM tori to a
special model, of great interest, whose relevance for the KAM problem
was pointed out by Thirring, [T] (see [G] for a short discussion of the
model). The idea of exposing Eliasson's method through simple particular
cases appears in [V], where results of the type of the ones discussed
here, and more general ones, are announced.

The connection between the methods of [E] and the tree expansions in the
renormalization group approaches to quantum field theory and many body
theory can be found also in [G[. The connection between the
tree expansions and the breakdown of invariant tori is discussed in
[PV].

The Thirring model is a system of rotators interacting via a potential.
It is described by the hamiltonian (see [G] for a motivation of the
name):
$$\fra12 J^{-1}\AA\cdot\AA\,+\,\e f(\aa)\Eq(1)$$
where $J$ is the (diagonal) matrix of the inertia moments,
$\AA=(A_1,\ldots,A_l)\in R^l$ are their angular momenta and
$\aa=(\a_1,\ldots,\a_l)\in T^l$ are the angles describing their
positions: the matrix $J$ will be supposed non singular; but we only
suppose that $\min_{j=1,\ldots,l}J_j=J_0>0$, and {\it no assumption} is
made on the size of the {\it twist rate} $T=\min J_j^{-1}$: the results
will be uniform in $T$ (hence the name ``twistless'': this is not a
contradiction with the necessity of a twist rate in the general
problems, see problems 1, 16, 17 in \S5.11 of [G2], and [G]).  We
suppose $f$ to be an even trigonometric polynomial of degree $N$:
$$f(\aa)=\sum_{0<|\nn|\le N} f_\nn\,\cos\nn\cdot\aa, \qquad
f_\nn=f_{-\nn}\Eq(2)$$
We shall consider a "rotation vector" $\oo_0=(\o_1,\ldots,\o_l)\in R^l$
verifying a {\it strong diophantine property} (see, however, the final
comments) with dophantine constants $C_0,\t,\g,c$; this means that:
$$\eqalign{
1)\kern1.truecm&
C_0|\oo_0\cdot\nn|\ge |\nn|^{-\t},\kern3.5cm\V0\ne\nn\in Z^l\cr
2)\kern1.truecm&
\min_{0\ge p\ge n}\big|C_0|\oo_0\cdot\nn|-\g^{p}\big|>\g^{n+1}\qquad
{\rm if}\ n\le0,\ 0<|\nn|\le (\g^{n+c})^{-\t^{-1}}\cr}
\Eq(3)$$
and it is easy to see that the {\it strongly diophantine vectors} have
full measure in $R^l$ if $\g>1$ and $c$ are fixed and if $\t$ is fixed
$\t>l-1$: we take $\g=2,c=3$ for simplicity; note that 2) is empty if
$n>-3$ or $p<n+3$.  We shall set $\AA_0=J\oo_0$.  And a special example
can be the model $f_0(\aa)=J_0\oo_0^2 (\cos\a_1+\cos(\a_1+\a_2))$.
\*

{\it We look for an $\e$--analytic family of motions starting at
$\aa=\V0$ and having the form:
$$\AA=\AA_0+ \V H(\oo_0t;\e),\qquad\aa=
\oo_0t+\V h(\oo_0 t;\e)\Eq(4)$$
with $\V H(\pps;\e),\V h(\pps;\e)$ analytic in $\pps\in T^l$ and in $\e$
close to $0$. We shall prove that such functions exist and are analytic
for $|\Im \psi_j|<\x$ for $|\e|<\e_0$ with:
$$\e_0^{-1}= b J_0^{-1}C_0^2f_0 N^{2+l} e^{c N}e^{\x N}\Eq(5)$$
where $b,c$ are $l$--dependent positive constants, $f_0=\max_\nn |f_\nn|$.
This means that the set $\AA=\AA_0+\V H(\pps;\e), \, \aa=\pps+\V
h(\pps;\e)$ described as $\pps$ varies in $T^l$ is, for $\e$ small
enough, an invariant torus for \equ(1), which is run quasi
periodically with angular velocity vector $\oo_0$. It is a family of
invariant tori coinciding, for $\e=0$, with the torus
$\AA=\AA_0,\,\aa=\pps\in T^l$. One recognizes a version of the KAM
theorem. The proof that follows simplifies the one reported
in [G].}
\*
Supposing $J_0\=J_1<J_2$ the uniformity in $J_2$ (\ie what we call the
{\it twistless} property) implies that the same $\e_0$ can be used as an
estimate of the radius of convergence in $\e$ of the power series
describing the KAM tori with rotation vector $\oo_0=(\o_1,\o_2)$ in the
system $(2J_1^{-1}A_1^2+\o_2 A_2+\e f_0(\a_1,\a_2)$, which is one of the
most studied hamiltonian systems. The estimate can be improved. Note
that a careful analysis of the proof of the KAM theorem also shows the
uniformity in the twist rate in the case \equ(1).

Calling $\V H^{(k)}(\pps),\V h^{(k)}(\pps)$ the $k$-th order
coefficients of the Taylor expansion of $\V H,\V h$ in powers of $\e$
and writing the equation of motion as $\dot\aa=\oo_0+J^{-1}\AA$ and
$\dot\AA=-\e\dpr_\aa f(\aa)$ we get immediately recursion
relations for $\V H^{(k)},\V h^{(k)}$. Namely
${\V\o}_0\cdot\V\dpr\,h^{(k)}_j=J^{-1}_j H^{(k)}_j$ and:
$${\V\o}_0\cdot\V\dpr\,H^{(k)}_j=-
\sum_{m_1,\ldots,m_l\atop|\mm|>0}\fra1{\prod_{s=1}^l m_s!}
\dpr_{\a_j}\, \dpr^{m_1+\ldots+m_l}_{\a_1^{m_1}\ldots\a_l^{m_l}} f(\oo_0
t)\cdot{\sum}^*\prod_{s=1}^l\prod_{j=1}^{m_s} h^{(k^s_j)}_s(\oo_0 t)\Eq(6)$$
where the $\sum^*$ denotes summation over the integers $k^s_j\ge1$
with: $\sum_{s=1}^l\sum_{j=1}^{m_s}k^s_j=k-1$.

The trigonometric polynomial $\V h^{(k)}(\pps)$ will be completely
determined (if possible at all) by requiring it to have $\V0$ average
over $\pps$, (note that $\V H^{(k)}$ has to have zero average over
$\pps$).  For $k=1$ one easily finds:
$$\tst\V h^{(1)}(\pps)=-\sum_{\nn\ne\V0}
\fra{J^{-1}\nn}{(i\oo_0\cdot\nn)^2}f_\nn\,e^{i\nn\cdot\pps}\Eq(7)$$
Suppose that $\V h^{(k)}(\pps)$ is a trigonometric polynomial of degree
$\le k N$, odd in $t$, for $1\le k< k_0$.  Then we see immediately that
the r.h.s.  of \equ(6) is odd in $t$.  This means that the r.h.s. of
\equ(6) has zero average in $t$, hence in $\pps$, and the second of
\equ(6) can be solved for $k=k_0$.  It yields an even function $\V
H^{(k_0)}(\pps)$ which is defined up to a constant which, however, must
be taken such that $\V H^{(k_0)}(\pps)$ has zero average, to make
$\oo\cdot\Dpr h^{(k)}_j=J_j^{-1} H^{(k)}_j$ soluble.  Hence the equation
for $\V h^{(k)}$ can be solved (because the r.h.s.  has zero average)
and its solution is a trigonometric polynomial in $\pps$, odd if $\V
h^{(k)}$ is determined by imposing that its average over $\pps$
vanishes.

Hence the \equ(7) provide an algorithm to evaluate a formal power series
solution to our problem. It has been remarked, [E],[V], see also [G],
that \equ(6) yields a {\it diagrammatic expansion} of $\V h^{(k)}$. We
simply "iterate" it until only $h^{(1)}$, given by \equ(7), appears.

Let $\th$ be a tree diagram: it will consist of a family of "lines" (\ie
segments) numbered from $1$ to $k$ arranged to form a (rooted) tree
diagram as in the figure:
\insertplot{240pt}{170pt}{
\ins{-35pt}{90pt}{\it root}
\ins{25pt}{110pt}{$j$}
\ins{60pt}{85pt}{$v_0$}
\ins{55pt}{115pt}{$\nn_{v_0}$}
\ins{115pt}{132pt}{$j_{1}$}
\ins{152pt}{120pt}{$v_1$}
\ins{145pt}{155pt}{$\nn_{v_1}$}
\ins{110pt}{50pt}{$v_2$}
\ins{190pt}{100pt}{$v_3$}
\ins{230pt}{160pt}{$v_5$}
\ins{230pt}{120pt}{$v_6$}
\ins{230pt}{85pt}{$v_7$}
\ins{230pt}{-10pt}{$v_{11}$}
\ins{230pt}{20pt}{$v_{10}$}
\ins{200pt}{65pt}{$v_4$}
\ins{230pt}{65pt}{$v_8$}
\ins{230pt}{45pt}{$v_9$}
}{gvnn}
\kern1.3cm
\didascalia{fig. 1: A tree diagram $\th$ with
$m_{v_0}=2,m_{v_1}=2,m_{v_2}=3,m_{v_3}=2,m_{v_4}=2$ and $m=12$,
$\prod m_v!=2^4\cdot6$, and some decorations. The line numbers,
distinguishing the lines, are not shown.}

To each vertex $v$ we attach a "mode label" $\nn_v\in Z^l,\,|\nn_v|\le N$
and to each branch leading to $v$ we attach a "branch label"
$j_v=1,\ldots,l$.  The order of the diagram will be $k=$ number of vertices
$=$ number of branches (the tree root will not be regarded as a vertex).

We imagine that all the diagram lines have the same length (even though
they are drawn with arbitrary length in fig.  1).  A group acts on the
set of diagrams, generated by the permutations of the subdiagrams having
the same vertex as root.  Two diagrams that can be superposed by the
action of a transformation of the group will be regarded as identical
(recall however that the diagram lines are numbered, \ie are regarded as
distinct, and the superpositon has to be such that all the decorations
of the diagram match).  Trees diagrams are regarded as partially ordered
sets of vertices (or lines) with a minimal element given by the root (or
the root line). We shall imagine that each branch carries also an arrow
pointing to the root (``gravity'' direction, opposite to the order).

We define the "momentum" entering $v$ as $\nn(v)=\sum_{w\ge v}\nn_w$. If
from a vertex $v$ emerge $m_1$ lines carrying a label $j=1$, $m_2$
lines carrying $j=2$, $\ldots$, it follows that \equ(6) can be
rewritten:
$$\V h^{(k)}_{\nn j}=\fra1{k!}
{\sum}^*\prod_{v\in\th}\fra{(i J^{-1}\nn_v)_{j_v}\,
f_{\nn_v}\prod_{s=1}^l(i\nn_v)^{m_s}_s}{(i\oo_0\cdot\nn(v))^2}
\Eq(8)$$
with the sum running over the diagrams $\th$ of order $k$ and with
$\nn(v_0)=\nn$; and the combinatorics can be checked from \equ(6), by
taking into account that we regard the diagram lines as all different
(to fix the factorials).

The ${}^*$ recalls that the diagram $\th$ can and will be supposed such
that $\nn(v)\ne\V0$ for all $v\in\th$ (by the above remarked parity
properties).  Note that \equ(8) is implied by the corresponding (6.23)
of [G]: one can check that the two formulae coincide (by summing over
what in [G] are called the ``fruit values'').  The theory in [G] is in
fact a little more general, although it is really applied to the same
Thirring model.

There are other diagrams, however, which we would like to eliminate.
They are the diagrams with nodes $v',v$, with $v'<v$, not necessarily
nearest neighbours, such that $\nn(v)=\nn(v')$. Such diagrams are,
according to Eliasson's terminology, {\it in resonance}.

To see this, {\it for the purpose of illustration}, we shall first
restrict the sum in \equ(8) to a sum over diagrams such that:
\item{I)} $\nn(v)\ne\V0$ \ if\ $v$ is any vertex.
\item{II)} $\nn(v)\ne\nn(v')$\ for all pairs of comparable vertices
$v',v$, (not necessarily next to each other in the diagram order,
however), with $v\ge v_0$.\*

There are at most $2^{2k}k!$ diagrams and the labels $j$ can be at most
$l^k$ while the $\nn_v$ can be chosen in a number of ways bounded by
$(2N+1)^{lk}<(3N)^{lk}$. Therefore, if $f_0=\max_\nn |f_\nn|$:
$$\ifnum\mgnf=1 \tst \else \relax \fi
|h^{(k)}_{\nn j}|\le N (3N)^{lk} 2^{2k}l^k \fra{f_0^k C_0^{2k}}{J_0^k}
N^{2k}\,
{\displaystyle\max_\th\prod_{v\in\th}}
(C_0\oo_0\cdot\nn(v))^{-2}\le
(f_0C_0^2J_0^{-1})^k N^{(l+2)k+1}(2l3^l)^k M\Eq(9)$$
where the maximum is over the diagrams $\th$ verifying the I),II) above.
Hence the whole problem is reduced to estimating the maximum with $M$.

Let $\z(q)\= (3 N^\t q^\t)^{-1}$, where $\t$ is the diophantine
constant in \equ(3): then $C_0|\oo_0\cdot\nn|\ge 3\z(q)$ if $0<|\nn|\le
q$: we can say that $\nn\in Z^l$ is "$q$--singular" if
$|\oo_0\cdot\nn|<\z(q)$.  Then the following (extension) of a lemma
by Brijuno, see [P], holds for diagrams of degree $k$ verifying I),II)
above.  Fixed $q\ge1$ let $N(k)$ be the number of $q$--resonant
harmonics, among the vertex momenta $\nn(v)$ of the vertices $v\in\th$.
Then $N(k)\le2\fra{k}{q}$.

Assuming the above claim true (we shall not use it outside the present
heuristic argument), we fix an exponentially decreasing sequence $\g^n$;
the choice $\g=2$ recommends itself.  The number of $q=\g^{-n}$--singular
harmonics which are not $q=\g^{-(n+1)}$--singular is bounded by $2k
\g^{-n}$, (being trivially bounded by the number of $\g^{-n}$--singular
harmonics!).  Hence:
$$\prod_{v\in \th}\fra1{(C_0\oo_0\cdot\nn(v))^2}\le \prod_{n\le-1}^{-\io}
\z(\g^{(n-1)})^{-4k \g^{n}}\le
\exp 4k\sum_{n\ge1}2^{-n}\log (3\cdot 2^{\t(n-1)}N^\t)=M\Eq(10)$$
therefore the series for the approximation to $\V h^{(k)}_\nn$, that we
are considering because of the extra restriction in the sum \equ(8),
has radius of convergence in $\e$ bounded below.

However {\it there are} resonant diagrams.  The key remark is that they
cancel {\it almost exactly}.  The reason is very simple: imagine to
detach from a diagram $\th$ the subdiagram $\th_2$ with first node $v$.
Then attach it to all the {\it remaining} vertices $w\ge v', w\in
\th/\th_2$.  We obtain a family of diagrams whose contributions to
$h^{(k)}$ differ because:
\item{1)} some of the branches above $v'$
have changed total momentum by the amount $\nn(v)$: this means that some
of the denominators $(\oo\cdot\nn(w))^{-2}$ have become
$(\oo\cdot\nn(v)\pm\e)^{-2}$ if $\e\=\oo_0\cdot\nn(v)$; and:
\item{2)} because there is one of the vertex factors which changes by
taking successively the values $\n_{wj}$, $j$ being the line label of
the line leading to $v$, and $w\in\th/\th_2$ is the vertex to which such
line is reattached.
\*

Hence if $\oo\cdot\nn=\e=0$ we would build in this {\it resummation}
a quantity proportional to: $\sum \nn_w= \nn(v)-\nn(v')$ which is zero,
because $\nn(v)=\nn(v')$ means that the sum of the $\nn_w$'s vanishes.
Since $\oo\cdot\nn=\e\ne0$ we can expect to see a sum of order $\e^2$,
if we sum as well on a overall change of sign of the $\nn_w$ values
(which sum up to $\V0$).

But this can be true only if $\e\ll \oo\cdot\nn'$, for any line momentum
$\nn'$ of a line in $\th/\th_2$.  If the latter property is not true
this means that $\oo\cdot\nn'$ is small and that there are many vertices
in $\th/\th_2$ of order of the amount needed to create a momentum with
small divisors of order $\e$.

Examining carefully the proof of Brjiuno's lemma one sees that such
extreme case would be essentially also treatable. Therefore the problem
is to show that the two regimes just envisaged (and their "combinations")
do exhaust all possibilities.

Such problems are very common in renormalization theory and are called
"overlapping divergences". Their systematic analysis is made through the
renormalization group methods. We argue here that Elliasson's method can
be interpreted in the same way.

The above introduced diagrams will play the role of {\it Feynman
diagrams}; and they will be plagued by overlapping divergences. They
will therefore be collected into another family of graphs,
that we shall call {\it trees}, on which the bounds are easy.
The $(\oo\cdot\nn)^{-2}$ are the {\it propagators}, in our analogy.

We fix an {\it scaling} parameter $\g$, which we take $\g=2$
for consistency with \equ(3), and we also define $\oo\=C_0\oo_0$: it
is an adimensional frequency.  Then we say that a propagator
$(\oo\cdot\nn)^{-2}$ is {\it on scale $n$} if $2^{n-1}<|\oo\cdot\nn|\le
2^n$, for $n\le0$, and we set $n=1$ if $1<|\oo\cdot\nn|$.

Proceeding as in quantum field theory, see [G3], given a diagram $\th$
we can attach a {\it scale label} to each line $v'v$ in \equ(8) (with
$v'$ being the vertex preceding $v$): it is
equal to $n$ if $n$ is the scale of the line propagator.
Note that the labels thus attached to a diagram are uniquely
determined by the diagram: they will have only the function of
helping to visualize the orders of magnitude of the various diagram
lines.

Looking at such labels we identify the connected clusters $T$ of
vertices that are linked by a continuous path of lines with the same
scale label $n_T$ or a higher one.  We shall say that {\it the cluster
$T$ has scale $n_T$}.

Among the clusters we consider the ones with the property that there is
only one diagram line entering them and only one exiting and both carry
the same momentum. Here we use that the diagram lines carry an arrow
pointing to the root: this gives a meaning to the words ``incoming'' and
``outgoing''.

If $V$ is one such cluster we denote $\l_V$ the incoming line: the line
scale $n=n_{\l_V}$ is smaller than the smallest scale $n'=n_V$ of the
lines inside $V$.  We call $w_1$ the vertex into which the line $\l_V$
ends, inside $V$.  {\it We say that such a $V$ is a {\it resonance} if
the number of lines contained in $V$ is $\le E\,2^{-n\e}$}, where
$n=n_{\l_V}$, and $E,\e$ are defined by:
$E\=2^{-3\e}N^{-1},\,\e=\t^{-1}$.  We shall say that $n_{\l_V}$ is {\it
the resonance scale}.

Let us consider a diagram $\th$ and its clusters.  We wish to estimate
the number $N_n$ of lines with scale $n\le0$ in it, assuming $N_n>0$.

Denoting $T$ a cluster of scale $n$ let $m_T$ be the number of
resonances of scale $n$ contained in $T$ (\ie with incoming lines of
scale $n$), we have the following inequality, valid for any diagram
$\th$:
$$N_n\le\fra{3k}{E\,2^{-\e n}}+\sum_{T, \,n_T=n}
(-1+m_T)\Eq(11)$$
with $E=N^{-1}2^{-3\e},\e=\t^{-1}$. This is a version of Brjuno's
lemma: a proof is in appendix.

Consider a diagram $\th^1$ we define the family $\FF(\th^1)$ generated
by $\th^1$ as follows. Given a resonance $V$ of $\th^1$ we detach the
part of $\th^1$ above $\l_V$ and attach it successively to the points
$w\in\tilde V$, where $\tilde V$ is the set of vertices of $V$
(including the endpoint $w_1$ of $\l_V$ contained in $V$) outside the
resonances contained in $V$. Note that all the lines $\l$ in $\tilde V$
have a scale $n_\l\ge n_V$.

For each resonance $V$ of $\th^1$ we shall call $M_V$ the number of
vertices in $\tilde V$.  To the just defined set of diagrams we add the
diagrams obtained by reversing simoultaneously the signs of the vertex
modes $\nn_w$, for $w\in \tilde V$: the change of sign is performed
independently for the various resonant clusters.  This defines a family
of $\prod 2M_V$ diagrams that we call $\FF(\th_1)$. The number
$\prod 2M_V$ will be bounded by $\exp\sum2M_V\le e^{2k}$.

It is important to note that the definition of resonance is such that
the above operation (of shift of the vertex to which the line entering
$V$ is attached) does not change too much the scales of the diagram
lines inside the resonances: the reason is simply that inside a
resonance of scale $n$ the number of lines is not very large being
$\le\lis N_n\=E\,2^{-n\e}$.

Let $\l$ be a line, in a cluster $T$, contained inside the resonances
$V=V_1\subset V_2\subset\ldots$ of scales $n=n_1>n_2>\ldots$; then the
shifting of the lines $\l_{V_i}$ can cause at most a change in the size
of the propagator of $\l$ by at most $2^{n_1}+2^{n_2}+\ldots< 2^{n+1}$.

Since the number of lines inside $V$ is smaller than $\lis N_n$ the
quantity $\oo\cdot\nn_\l$ of $\l$ has the form
$\oo\cdot\nn^0_\l+\s_\l\oo\cdot\nn_{\l_V}$ if $\nn^0_\l$ is the momentum
of the line $\l$ "inside the resonance $V$", \ie it is the sum of all
the vertex modes of the vertices preceding $\l$ in the sense of the
line arrows, but contained in $V$; and $\s_\l=0,\pm1$.

Therefore not only $|\oo\cdot\nn^0_\l|\ge 2^{n+3}$ (because $\nn^0_\l$
is a sum  of $\le \lis N_n$ vertex modes, so that $|\nn^0_\l|\le N\lis
N_n$) but $\oo\cdot\nn^0_\l$ is "in the middle" of the diadic interval
containing it and by \equ(3) does not get out of it if we add a quantity
bounded by $2^{n+1}$ (like $\s_\l\oo\cdot\nn_{\l_V}$). Hence no line
changes scale as $\th$ varies in $\FF(\th^1)$, if $\oo_0$ verifies \equ(3).

{\it This implies, by the strong diophantine hypothesis on $\oo_0$,
\equ(3), that the resonant clusters of the diagrams in $\FF(\th^1)$ all
contain the same sets of lines, and the same lines go in or out of each
resonance (although they are attached to generally distinct vertices
inside the resonances: the identity of the lines is here defined by the
number label that each of them carries in $\th^1$).  Furthermore the
resonance scales and the scales of the resonant clusters, and of all the
lines, do not change.}

Let $\th^2$ be a diagram not in $\FF(\th^1)$ and construct
$\FF(\th^2)$, \etc.  We define a collection
$\{\FF(\th^i)\}_{i=1,2,\ldots}$ of pairwise disjoint families of
diagrams.  We shall sum all the contributions to $\V h^{(k)}$ coming
from the individual members of each family.  This is the {\it
Eliasson's resummation}.

We call $\e_V$ the quantity $\oo\cdot\nn_{\l_V}$ associated with the
resonance $V$. If $\l$ is a line with both extremes in $\tilde V$ we can
imagine to write the quantity $\oo\cdot\nn_\l$ as
$\oo\cdot\nn^0_\l+\s_\l\e_V$, with $\s_\l=0,\pm1$. Since $|\oo\cdot\nn_\l|>
2^{n_V-1}$ we see that the product of the propagators is holomorphic in
$\e_V$ for $|\e_V|<2^{n_V-3}$.\ \footnote{${}^2$}{
In fact $|\oo\cdot\nn^0_\l|\ge 2^{n+3}$ because $V$ is a resonance;
therefore $|\oo\cdot\nn_\l|\ge 2^{n+3}-2^{n+1}\ge 2^{n+2}$ so that $n_V\ge
n+3$. On the other hand note that $|\oo\cdot\nn^0_\l|> 2^{n_V-1}-2^{n+1}$ so
that $|\oo\cdot\nn_\l^0+\s_\l\e_V|\ge 2^{n_V-1}-2^{n+1}-2^{n_V-3}\ge
2^{n_V-1}-2^{n_V-3}\ge 2^{n_V-2}$, for $|\e_V|< 2^{n_V-3}$.}
While $\e_V$ varies in such complex disk
the quantity $|\oo\cdot\nn_\l|$ does not become smaller than
$2^{n_V-1}- 2\,2^{n_V-3}\ge2^{n_V-2}$. Note the main point here: the
quantity $2^{n_V-3}$ will usually be $\gg 2^{n_{\l_V}-1}$ which is the
value $\e_V$ actually can reach in every diagram in $\FF(\th^1)$; this
can be exploited in applying the maximum priciple, as done below.

It follows that, calling $n_\l$ the scale of the line $\l$ in $\th^1$,
each of the $\prod 2 M_V\le e^{2k}$ products of propagators
of the members of the family $\FF(\th^1)$ can be bounded above by
$\prod_\l\,2^{-2(n_\l-2)}=2^{4k}\prod_\l\,2^{-2n_\l}$, if regarded as a
function of the quantities $\e_V=\oo\cdot\nn_{\l_V}$, for $|\e_V|\le
\,2^{n_V-3}$, associated with the resonant clusters $V$.  This even
holds if the $\e_V$ are regarded as independent complex parameters.

By construction it is clear that the sum of the $\prod 2M_V\le e^{2k}$
terms, giving the contribution to $\V h^{(k)}$ from the trees in
$\FF(\th^1)$, vanishes to second order in the $\e_V$ parameters (by the
approximate cancellation discussed above).  Hence by the maximum
principle, and recalling that each of the scalar products in \equ(8) can
be bounded by $N^2$, we can bound the contribution from the family
$\FF(\th^1)$ by:
$$\left[\fra1{k!} N\Big(\fra{f_0 C_0^2 N^2}{J_0}\Big)^k 2^{4k} e^{2k}
\prod_{n\le0}2^{-2nN_n}\right]\left[\prod_{n\le0}\prod_{T,\,n_T=n}
\prod_{i=1}^{m_T}\,2^{2(n-n_{i}+3)}\right]\Eq(12)$$
where:
\acapo
1) $N_n$ is the number of propagators of scale $n$ in $\th^1$ ($n=1$
does not appear as $|\oo\cdot\nn|\ge1$ in such cases),\acapo
2) the first square bracket is the bound on the product of
individual elements in the family $\FF(\th^1)$ times the bound $e^{2k}$
on their number,
\acapo
3) The second term is the part coming from the maximum principle, applied
to bound the resummations, and is explained as follows.
\acapo
i) the dependence on the variables $\e_{V_i}\=\e_i$ relative to
resonances $V_i\subset T$ with scale $n_{\l_V}=n$ is holomorphic for for
$|\e_i|<\,2^{ n_i-3}$ if $n_i\=n_{V_i}$, provided $n_i>n+3$ (see above).
\acapo
ii) the resummation says that the dependence on the $\e_i$'s has a
second order zero in each.  Hence the maximum principle tells us that
we can improve the bound given by the first factor in \equ(12) by the
product of factors $(|\e_i|\,2^{-n_i+3})^2$ if $n_i>n+3$.  If $ n_i\le
n+3$ we cannot gain anything: but since the contribution to the bound
from such terms in \equ(12) is $>1$ we can leave them in it to simplify
the notation, (of course this means that the gain factor can be
important only when $\ll1$).

Hence substituting \equ(11) into \equ(12) we see that the $m_T$ is taken
away by the first factor in $\,2^{2n}2^{-2n_{i}}$, while the remaining
$\,2^{-2n_i}$ are compensated by the $-1$ before the $+m_T$ in \equ(11),
taken from the factors with $T=V_i$,  (note that there are always enough
$-1$'s).

Hence the product \equ(12) is bounded by:
$$\fra1{k!}N\,(C_0^2J_0^{-1}f_0 N^2)^k e^{2k}2^{4k}2^{6k}
\prod_n\,2^{-6 n k E^{-1}\,2^{\e n}}\le \fra1{k!}N\, B_0^k\Eq(13)$$
with: $B_0=2^{10}e^2 C_0^2 f_0 N^2 \exp [N\,
(2^{2+2\t^{-1}}\log2\big)\sum_{p=1}^\io p 2^{-p\t^{-1}}]$.

To sum over the trees we note that, fixed $\th$ the collection of
clusters is fixed.  Therefore we only have to multiply \equ(13) by the
number of diagram shapes for $\th$, ($\le 2^{2k}k!$), by the number of
ways of attaching mode labels, ($\le (3N)^{lk}$), so that we can bound
$|h^{(k)}_{\nn j}|$ by \equ(5).
\*
\0{\it Comments.}
\*
The strong diophantine condition is quite unpleasant as
it seems to put an extra requirement on $\oo_0$: I think that in fact
such condition is not necessary. Given $\oo_0$ verifying the first of
\equ(3) with some constant $\bar C_0$ and some $\t$ we can imagine to
define $C_0\=2^\t \bar C_0$: this leaves the first of \equ(3) still valid.
One should then note that the $0<|\nn|\le(2^{n+3})^{-\t^{-1}}$
will imply that if $0<|\nn|\le(2^{n+3})^{-\t^{-1}}$
the numbers $|\oo\cdot\nn|$ are spaced by at least
$(2(2^\t 2^{n+3})^{-\t^{-1}})^{-\t}=2^{n+3}$. Hence we can find a sequence
$\g_n$ such that $1\le\g_n 2^{-n}\le2$ and
$||\oo\cdot\nn|-\g_p|\ge 2^{n+2}$ for $0\ge p\ge n$. Defining
a propagator $x^{-2}$ to have "scale $n$" if $\g_{n-1}<|x|\le\g_n$ it
should be possible to perform the proof without the second assumption
in \equ(3).  This will be interesting to check if it really works as it
would be fully constructive, given $\oo_0$ verifying the first of
\equ(3), and as it would cover the most studied cases like the case
$l=2$ and $\oo=(\o_1,\o_2)=(1,\o)$ with $\o$ a quadratic irrational, or
a number with uniformly bounded continued fraction entries. It has also
the conceptual advantage that the sequence of scales $\g_n$ is not
``prescribed a priori'' but it is determined by the arithmetic
properties of the rotation vector $\oo_0$.

\penalty-200\vskip1.truecm
{\bf Appendix A1: Resonant Siegel-Brjuno bound.}
\penalty10000\vskip0.5truecm

Calling $N^*_n$ the number of non resonant lines carrying a scale label
$\le n$. We shall prove first that $N^*_n\le 2k (E 2^{-\e n})^{-1}$ if
$N^*_n>0$.

If $\th$ has the root line with scale $>n$ then calling
$\th_1,\th_2,\ldots,\th_m$ the subdiagrams of $\th$ emerging from the
first vertex of $\th$ and with $k_j>E\,2^{-\e n}$ lines, it is
$N^*(\th)=N^*(\th_1)+\ldots+N^*(\th_m)$ and the statement is inductively
implied from its validity for $k'<k$ provided it is true that
$N^*(\th)=0$ if $k<E2^{-\e n}$, which is is certainly the case if $E$ is
chosen as in \equ(11).\footnote{${}^3$}{\nota Note that if $k\le
E\,2^{-n\e}$ it is, for all momenta $\nn$ of the lines, $|\nn|\le N E
\,2^{-n\e}$, \ie $|\oo\cdot\nn|\ge(NE\,2^{-n\e})^{-\t}=2^3\,2^{n}$ so
that there are {\it no} clusters $T$ with $n_T=n$ and $N^*=0$. The
choice $E=N^{-1}2^{-3\e}$ is convenient: but this, as well as the whole
lemma, remains true if $3$ is replaced by any number larger than $1$.
The choice of $3$ is made only to simplify some of the arguments based
on the resonance concept.}

In the other case it is $N^*_n\le 1+\sum_{i=1}^mN^*(\th_i)$, and if
$m=0$ the statement is trivial, or if $m\ge2$ the statement is again
inductively implied by its validity for $k'<k$.

If $m=1$ we once more have a trivial case unless the order $k_1$ of
$\th_1$ is $k_1>k-\fra12 E\,2^{-n\e}$.  Finally, and this is the real
problem as the analysis of a few examples shows, we claim that in the
latter case the root line is either a resonance or it has scale $>n$.

Accepting the last statement it will be: $N^*(\th)=1+N^*(\th_1)=
1+N^*(\th'_1)+\ldots+N^*(\th'_{m'})$, with $\th'_j$ being the $m'$
subdiagrams emerging from the first node of $\th'_1$ with orders
$k'_j>E\,2^{-\e n}$: this is so because the root line of $\th_1$ will
not contribute its unit to $N^*(\th_1)$.  Going once more through the
analysis the only non trivial case is if $m'=1$ and in that case
$N(\th'_1)=N^*(\th"_1)+\ldots+N(\th"_{m"})$, \etc, until we reach a
trivial case or a diagram of order $\le k-\fra12 E\,2^{-n\e}$.

It remains to check that if $k_1>k-\fra12E\,2^{-n\e}$ then the root line of
$\th_1$ has scale $>n$, unless it is entering a resonance.

Suppose that the root line of $\th_1$ is not entering a resonance. Note
that $|\oo\cdot\nn(v_0)|\le\,2^n,|\oo\cdot\nn(v)|\le
\,2^n$, if $v_0,v_1$ are the first vertices of $\th$ and $\th_1$
respectively.  Hence $\d\=|(\oo\cdot(\nn(v_0)-\nn(v_1))|\le2\,2^n$ and
the diophantine assumption implies that $|\nn(v_0)-\nn(v_1)|>
(2\,2^n)^{-\t^{-1}}$, or $\nn(v_0)=\nn(v_1)$.  The latter case being
discarded as $k-k_1<\fra12E\,2^{-n\e}$ (and we are not considering the
resonances), it follows that $k-k_1<\fra12E\,2^{-n\e}$ is inconsistent:
it would in fact imply that $\nn(v_0)-\nn(v_1)$ is a sum of $k-k_1$
vertex modes and therefore $|\nn(v_0)-\nn(v_1)|< \fra12NE\,2^{-n\e}$
hence $\d>2^3\,2^n$ which is contradictory with the above opposite
inequality.

A similar, far easier, induction can be used to prove that if $N^*_n>0$
then the number $p$ of clusters of scale $n$ verifies the bound
$p< k \,(E2^{-\e n})^{-1}-1$. Thus \equ(11) is proved.

{\it Remark}: the above argument is a minor adaptation of Brjiuno's
proof of Siegel's theorem, as remarkably exposed by P\"oschel, [P].

\penalty-200
\*
{\it Acknowledgements: I am indebted to J.  Bricmont and M.  Vittot for
many clarifying discussions; and in particular to S. Miracle-Sol\'e for
the same reasons and for his interest and suggestions..}
\*

\vskip0.5truecm

\penalty-200

{\bf References}

\vskip0.5truecm

\penalty10000

\item{[A] } Arnold, V.: {\it Proof of a A.N. Kolmogorov theorem on
conservation of conditionally periodic motions under small perturbations
of the hamiltonian function}, Uspeki Matematicheskii Nauk, 18, 13-- 40,
1963.

\item{[BG] } Benfatto, G., Gallavotti, G.: {\it Perturbation theory of
the Fermi surface in a quantum liquid. A general quasi particle
formalism and one dimensional systems}, Journal of Statistical Physics,
59, 541- 664, 1990. Benfatto, G., Gallavotti, G.: {\it Renormalization
group application tot he theory of the Fermi surface}, Physical
Review, B, 42, 9967-- 9972, (1990).

\item{[B] } Brjuno, A.: {\it The analytic form of differential
equations}, I: Transactions of the Moscow Mathematical Society, 25,
131-- 288, 1971; and II: 26, 199--239, 1972. And

\item{[E] } Eliasson L. H.: {\it Hamiltonian systems with linear normal
form near an invariant torus}, ed. G. Turchetti, Bologna Conference,
30/5 to 3/6 1988, World Scientific, 1989. And {\it Generalization of an
estimate of small divisors by Siegel}, ed. E. Zehnder, P. Rabinowitz,
book in honor of J. Moser, Academic press, 1990. But mainly:
{\it Absolutely convergent series expansions
for quasi--periodic motions}, report 2--88, Dept. of Math., University of
Stockholm, 1988.

\item{[G] } Gallavotti, G.: {\it Twistless KAM tori, quasi flat
homoclinic intersections, and other cancellations in the perturbation
series of certain completely integrable hamiltonian systems.  A
review.}, deposited in the archive {\tt mp\_arc@math.utexas.edu}.

\item{[G2] } Gallavotti, G.: {\it The elements of Mechanics}, Springer, 1983.

\item{[G3] } Gallavotti, G.: {\it Renormalization theory and ultraviolet
stability for scalar fields via renormalization group methods}, Reviews
in Modern Physics, 57, 471- 572, 1985. See also, Gallavotti, G.:
{\it Quasi integrable mechanical systems}, Les Houches, XLIII (1984),
vol. II, p. 539-- 624, Ed. K. Osterwalder, R. Stora, North Holland, 1986.

\item{[K] } Kolmogorov, N.: {\it On the preservation of conditionally
periodic motions}, Doklady Aka\-de\-mia Nauk SSSR, 96, 527-- 530, 1954. See
also: Benettin, G., Galgani, L., Giorgilli, A., Strelcyn, J.M.:
{\it A proof of Kolmogorov theorem on invariant tori using canonical
transormations defined by the Lie method}, Nuovo Cimento, 79B, 201--
223, 1984.

\item{[M] } Moser, J.: {\it On invariant curves of an area preserving
mapping of the annulus}, Nach\-rich\-ten Akademie Wiss. G\"ottingen, 11,
1-- 20, 1962.

\item{[P] } P\"oschel, J.: {\it Invariant manifolds of complex analytic
mappings}, Les Houches, XLIII (1984), vol. II, p. 949-- 964, Ed. K.
Osterwalder, R. Stora, North Holland, 1986.

\item{[PV] } Percival, I., Vivaldi, F.: {\it Critical dynamics and
diagrams}, Physica, D33, 304-- 313, 1988.

\item{[S] } Siegel, K.: {\it Iterations of analytic functions}, Annals
of Mathematics, {\bf 43}, 607-- 612, 1943.

\penalty10000

\item{[T] } Thirring, W.: {\it Course in Mathematical Physics}, vol. 1,
p. 133, Springer, Wien, 1983.

\item{[V]} Vittot, M.: {\it Lindstedt perturbation series in hamiltonian
mechanics: explicit formulation via a multidimensional Burmann--Lagrange
formula}, Preprint CNRS-- Luminy, case 907, F-13288, Marseille 1992.

\ciao